\documentstyle[prd,aps,epsf]{revtex}

\draft
\begin{document}
\twocolumn[\hsize\textwidth\columnwidth\hsize\csname
@twocolumnfalse\endcsname

\title{Uniqueness of (dilatonic) charged black holes and black $p$-branes in higher dimensions}

\author{${}^{(a)}$Gary W. Gibbons, ${}^{(b)}$Daisuke Ida and
${}^{(b),(c)}$Tetsuya Shiromizu}

\address{${}^{(a)}$DAMTP, Centre for Mathematical Sciences, The University of Cambridge,\\
Wilberforce Road, Cambridge CB3 0WA, United Kingdom \\
${}^{(b)}$Department of Physics, Tokyo Institute of Technology,
Tokyo 152-8551, Japan \\
${}^{(c)}$Advanced Research Institute for Science and Engineering,
Waseda University, Tokyo 169-8555, Japan}

\maketitle

\begin{abstract}
We prove the uniqueness of higher 
dimensional (dilatonic) charged black holes in  static and 
asymptotically flat spacetimes for arbitrary vector-dilaton
coupling constant. An  application to the uniqueness
of  a wide class of black p-branes
is also given.
\end{abstract}

\pacs{PACS: 04.50.+h; 04.70.Bw\qquad
DAMTP-2002-74;
TIT/HEP-480
}

\vskip2pc]

\vskip1cm

\section{Introduction}

The aim of  this paper is to  extend our recent work 
in which we generalized the four-dimensional Israel's theorems 
on the uniqueness of 
of static vacuum and electrically charged  black holes \cite{EM1} 
to higher dimensions \cite{Ida1},  \cite {Ida2}. We also gave 
a uniqueness theorem for a certain class of charged 
dilatonic black holes in higher dimensions  \cite{Ida2}. 
There is also earlier work by Hwang  \cite{Hwang} on the vacuum case. 
The motivation for treating charged dilatonic black holes comes from 
string theory where  gauge fields often play an essential role. 
It is clearly important to be able deal with the most general 
vector-dilaton coupling constant and that is what we shall do in this 
present paper.

Another important motivation for our work is to continue our
programme of proving the uniqueness of static black p-brane solutions. 
Such solutions are  $(n+p)$-dimensional  spacetimes invariant 
under the action of a 
$p$-dimensional Abelian translation group. Reduction to $n$ spacetime 
dimensions produces a black hole solution of gravity coupled 
to one or more scalars and an electric $2$-form  or dual 
magnetic  $(n-2)$-form field strength. 
A uniqueness theorem for a black hole solution in $n$ spacetime 
dimensions which asserts the 
necessity of $SO(n-1)$ invariance acting on $S^n$ orbits 
is equivalent to asserting the invariance of the static black p-brane 
under the action of $SO(n-1)$ on the $n-1$ space transverse to the 
p-brane. As in the case of 4-dimensional black holes 
so with higher dimensional charged black holes and branes 
although we  expect that  a uniqueness theorem holds in the 
extreme case case we cannot expect such a theorem 
to assert $SO(n-1)$-invariance because of the existence 
of multi-p-brane solutions in static equilibrium. 
In this paper  we will only treat the non-extreme case, leaving the 
extreme case for a later date. We assume only electrically 
charged black holes coupled to a single 2-form and with a single 
scalar field. By duality we could also consider 
magnetically charged black hole coupled to an $(n-2)$ form. 
In previous work we were limited to the vacuum case\cite{Ida1} 
or to special values of the the vector-dilaton couplings constant\cite{Ida2}, 
which meant that our results only held for the cases $(n,p)=(5,3)$ or 
$(6,2)$. The results of this paper will remove this restriction:rather 
than generalizing to higher dimensions the argument given in \cite{EMD} 
as was done in our earlier paper \cite{Ida2}, we shall 
generalize the method introduced in four dimensions in  \cite{Simon}

As in the case of higher dimensional vacuum black holes, so 
in this case, it is essential to assume strict asymptotic flatness 
because analogues of the Bohm black holes \cite{Ida1,Ida2} also exist 
in the electrically charged case. In addition, just as in 3+1 dimensions, 
we also have to assume that the surface 
gravity is non-zero(non-extreme), otherwise one has multi-black holes 
solutions 
which generalize the Majumadar-Papapetrou (MP) solutions\cite{MP}. 
See Ref. \cite{MP2} for the discussion the uniqueness of the MP solution 
under some additional assumptions. 
We shall also prove the uniqueness of Gibbons-Maeda solution 
\cite{GM} of  the Einstein-Maxwell-dilaton system with the 
general dilaton coupling to the Maxwell field. This amounts to 
a  non-trivial generalisation of the earlier 
papers \cite{Ida1,Ida2,EMD,Simon}.

The rest of this paper is organised as follows. In the section two, 
we will prove the uniqueness of the higher dimensional 
Reissner-Nordstrom solution\cite{Tangherlini} among 
static black holes in the Einstein-Maxwell system. 
Then, in the section three, a black hole uniqueness for the most general 
Einstein-Maxwell-dilaton system is given following the four-dimensional 
proof in \cite{Simon}. In the section four, we address the uniqueness 
of black $p$-branes. Finally  in the section five we provide a summary 
of the results of the paper.

%The appendix A contains the description of the conformal positive
%energy theorem\cite{CPET} used in the section three.

\section{Charged Black Holes}

In this section, we consider the $n$-dimensional 
Einstein-Maxwell system given by the Lagrangian
\begin{equation}
L={}^{n}R-F^2,
\end{equation}
where $F$ is the Maxwell field, 
and prove the uniqueness of the static and 
electrically charged black holes in higher dimensions.

In general, the metric of an  $n$-dimensional 
static space-time has the form
%========<Equation>========%
%
\begin{eqnarray}
ds^2=-V^2dt^2+g_{ij}dx^idx^j,\label{metric}
\end{eqnarray}
%
%==========================%
where $V$ and $g_{ij}$ are independent of $t$ 
and they are regarded as quantities on the $t={\rm const.}$ 
hypersurface $\Sigma$. The event horizon $H$ is a Killing horizon 
located at the level set $V=0$, which is assumed to be non-degenerate. 
Then the static field equations become
%========<Equation>========%
%
\begin{eqnarray}
\nabla^2V&=&{C^2\over V}(\nabla\psi)^2, \label{einstein1}\\
\nabla^2\psi&=&{\nabla\psi\cdot\nabla V\over V},
\label{Maxwell}
\end{eqnarray}
and
\begin{eqnarray}
R_{ij}&=&{\nabla_i\nabla_j V\over V}
-{2\over V^2}\nabla_i\psi\nabla_j\psi
+{2(\nabla\psi)^2\over (n-2)V^2}g_{ij}
, \label{einstein2}
\end{eqnarray}
%
%==========================%
where $C=[2(n-2)/(n-3)]^{1/2}$, $\nabla$ and $R_{ij}$ denote 
covariant derivative and the Ricci tensor defined on 
$(\Sigma, g_{ij})$, respectively, and $\psi$ is the electrostatic 
potential such that $F=d\psi\wedge dt$. 

In asymptotically flat space-times, one can find an appropriate 
coordinate system in which the metric and electrostatic potential 
have  asymptotic expansions of the form
\begin{eqnarray}
V &=& 1-\frac{\mu}{r^{n-3}}+O(1/r^{n-2}),\\
g_{ij}&=&\left(
1+\frac{2}{n-3}\frac{\mu}{r^{n-3}}\right)\delta_{ij}+O(1/r^{n-2}),\\
\psi&=&{Q/C\over  r^{n-3}}+O(1/r^{n-2}),
\end{eqnarray}
respectively, where $\mu$, $Q={\rm const.}$ represent the ADM mass 
and the electric charge (up to  constant factors), respectively, 
and $r:=\sqrt{\sum_i(x^i)^2}$. We assume the non-extremal condition 
$\mu>|Q|$.

Consider the following two conformal transformations
\begin{eqnarray}
\hat g_{ij}^{\pm}=\Omega_{\pm}^2 g_{ij},
\end{eqnarray}
where
\begin{eqnarray}
\Omega_\pm^2=\left[\left({1\pm V\over 2}\right)^2-
{C^2\over 4}\psi^2\right]^{1/2}
\end{eqnarray}
Then we have two manifolds ($ \Sigma^{\pm}, \hat g_{ij}^{\pm}$).
On $\Sigma^{+}$, the asymptotic behavior of the metric
becomes
\begin{eqnarray}
\hat g_{ij}^{+} =\delta_{ij}+ O\left(1/r^{n-2} \right).\label{asymptotic}
%~~~(r\rightarrow +\infty).
\end{eqnarray}
On  $ \Sigma^{-}$, we have
\begin{eqnarray}
\hat g_{ij}^{-}& = &
%\left({\mu^2-Q^2\over2\mu}\right)^{4/(n-3)}}{r^{-4}\left(dr^2+r^2d\Omega^2_{n-2}\right)+O(1/r^5)\nonumber\\
%&& ~~~(r\rightarrow +\infty)\nonumber \\
%& = &
{
\left[(\mu^2-Q^2)/4\right]^{2/(n-3)}\over r^4}\delta_{ij}+O(1/r^5)
\nonumber\\
&=&\left[(\mu^2-Q^2)/4\right]^{2/(n-3)}
\left(d\varrho^2+\varrho^2d\Omega^2_{n-2}\right)
%\nonumber\\
%&&{}
+O(\varrho^5).\nonumber\\
%&&~~~(R\rightarrow 0).
\end{eqnarray}
where $d\Omega^2_{n-2}$ denotes the round sphere metric and $\varrho:=1/r$ 
has been defined. Pasting $( \Sigma^{\pm}, \hat g_{ij}^{\pm})$ across the
level set $V=0$ and adding a point $\{p\}$ at $\varrho=0$, 
we can construct a complete regular surface 
$\hat\Sigma= \Sigma^{+} \cup \Sigma^{-}\cup \{p\}$. The Ricci curvature 
$\hat R$ on $\Sigma^\pm$ becomes
\begin{eqnarray}
\Omega_\pm^2\hat R&=&\frac{\hat\nabla^2V}{V}
+\frac{2}{n-2}\frac{(\hat\nabla\psi)^2}{V^2}
-2(n-2)\frac{\hat\nabla^2\Omega_\pm}{\Omega_\pm}\nonumber\\
&&{}-(n-2)(n-5)\frac{(\hat\nabla\Omega_\pm)^2}{\Omega_\pm^2}
\nonumber\\
&=&\frac{1}{8V^2\Omega_\pm^{2(n-3)}}
\Biggl|2V\psi \hat\nabla V-(V^2-1+C^2\psi^2)\hat\nabla\psi\Biggr|^2
\nonumber\\
\label{ricci}
\end{eqnarray}
where we have used the identities
\begin{equation}
\hat\nabla^2(V\pm C\psi)=\pm \frac{C}{V}\hat\nabla\psi\cdot\hat\nabla(V\pm C\psi).
\end{equation}
Then the Ricci curvature on $\hat\Sigma$ is nonnegative.
Furthermore, Eq.~(\ref{asymptotic}) implies that
the total mass also vanishes on $\hat\Sigma$.
As a consequence of the positive mass theorem \cite{PET,Marika},
such a surface $\hat\Sigma$ must be flat
and
\begin{eqnarray}
2V\psi \hat\nabla V=(V^2-1+C^2\psi^2)\hat\nabla\psi
\end{eqnarray}
holds, which implies that the level surfaces of $V$ and $\psi$ coincide. 
In other words, the physical Cauchy surface $\Sigma$ is conformally flat. 
We shall now demonstrate that the conformally transformed event horizon 
$\hat H$ is a geometric sphere in $\hat\Sigma$. 
We choose $V$ as a local coordinate in a neighbourhood $U \subset \Sigma$. 
Let $\{x^A\}$ be coordinates on level sets of $V$ 
such that their trajectories are orthogonal to each level set. 
Then, the metric on $\Sigma$ can be  written in the form
\begin{equation}
g=\rho^2dV^2+h_{AB}dx^Adx^B,
\end{equation}
where $\rho^2:=(\nabla V)^2$.
Since $\Sigma$ is conformally flat, the Riemann invariant has 
a simple expression in this coordinate system: 
\begin{eqnarray}
{}^nR_{IJKL}{}^nR^{IJKL}&=&R_{ijkl}R^{ijkl}+4R_{0i0j}R^{0i0j}\nonumber\\
&  = & \frac{4(n-2)}{(n-3)V^2\rho^2}\left[k_{AB}k^{AB}+k^2\right.\nonumber\\
&&\left.{}+2 {\cal D}_A \rho
{\cal D}^A \rho\right],
\label{kretchmann}
\end{eqnarray}
where ${\cal D}_A$ denotes the covariant derivative on each level set of
$V$, and $k_{AB}$ is the second fundamental form of the level set.

The requirement that the event horizon $H$ is a regular surface
leads to the condition
\begin{eqnarray}
&&k_{AB}|_H=0,\\
&&{\cal D}_A\rho|_H=0.
\end{eqnarray}
In particular, $H$ is a totally geodesic surface in $\Sigma$.

Let us consider how the event horizon 
is embedded into the base space $(\hat\Sigma,\delta_{ij})$. 
Define the smooth function
\begin{equation}
v:=\left(1+V-C\psi\right)^{-1},
\end{equation}
which is the harmonic function on $(\hat\Sigma,\delta_{ij})$: $\nabla_0^2v=0$. 
In terms of this, we can 
adopt the following local expression for the flat space
\begin{eqnarray}
\delta_{ij}dx^idx^j=\hat \rho^2 dv^2+\hat h_{AB}dx^A dx^B.
\end{eqnarray}
The event horizon is located at some $v={\rm const.}$ surface $\hat H$.
The extrinsic curvature $\hat k_{AB}$ of the level set $v={\rm const.}$ 
can be expressed as
\begin{eqnarray}
\hat k_{AB}&=&\Omega_+k_{AB}+{1\over\hat\rho}
{\partial\Omega_+\over\partial v}h_{AB}.
\end{eqnarray}
Thus we obtain
\begin{equation}
\hat k_{AB}={1\over\hat\rho}{\partial\Omega_+\over\partial v}\Biggr|_H \hat h_{AB}.
\end{equation}
on $\hat H$. In other words, the embedding of $\hat H$ into the 
Euclidean $(n-1)$-space is totally umbilical. 
It is known that such a embedding must be hyperspherical \cite{sphere}, 
namely 
each connected component of $\hat H$ is a geometric sphere with a certain 
radius. The embedding of a hypersphere into the Euclidean space is known 
to be rigid\cite{rigid}, 
which means that we can always locate one connected component of 
$\hat H$ at the $r=r_0$ surface of $\tilde \Sigma$ 
without loss of generality. If there is only a single horizon, we have 
a boundary value problem for the Laplace equation $\nabla_0^2 v=0$ 
on the base space $\Omega:=E^{n-1}\setminus B^{n-1}$ 
with the Dirichlet boundary conditions. 
Such a solution must be spherically symmetric, so that the 
Birkhoff theorem implies that it is given by 
the Reissner-Nordstrom solutions.

%%%%%%%%%%%%%%%%%%%%%%%%%%%%%%%%%%

One may remove the assumption of the single horizon as follows. 
Consider the evolution of the level surface in 
Euclidean space. From the Gauss equation in Euclidean space one 
obtains  the evolution equation for the shear $\hat \sigma_{AB}:=
\hat k_{AB}-\hat k \hat h_{AB}/(n-2)$:
\begin{eqnarray}
\mbox \pounds_{\hat n}  \hat \sigma_{AB}
& = & \hat \sigma_A{}^C \hat \sigma_{CB}
+\frac{1}{n-2}\hat h_{AB} \hat \sigma_{CD} \hat \sigma^{CD} \nonumber
\\
& & -\frac{1}{\hat \rho}\biggl( {\hat {\cal D}}_A {\hat {\cal D}}_B
-\frac{1}{n-2}\hat h_{AB} {\hat {\cal D}}^2  \biggr)\hat \rho,
\end{eqnarray}
where $\hat n$ denotes the unit normal to the level set of $v$. 
Using $\nabla_0^2 v=0$, we obtain
%we can derive the equation for
%${\tilde {\cal D}}_A {\rm ln}\tilde \rho$ as
\begin{eqnarray}
\mbox \pounds_{\hat n} {\hat {\cal D}}_A {\rm ln}\hat \rho&=&
\hat k {\hat {\cal D}}_A {\rm ln}\hat \rho +{\hat {\cal D}}_A \hat 
k,\\
\mbox \pounds_{\hat n} \hat k &=& -
||\hat\sigma||^2-\frac{1}{n-2}k^2-\frac{1}{\hat \rho}
{\hat {\cal D}}^2 \hat \rho,\\
\mbox \pounds_{\hat n} {\cal D}_A \hat k &=&{\hat {\cal D}}_A \mbox
\pounds_{\hat n}
\hat k + ({\hat {\cal D}}_A {\rm ln}\hat \rho)(  \mbox \pounds_{\hat n}
\hat k).
\end{eqnarray}
From the above equations, it can be seen that
\begin{eqnarray}
\hat \sigma_{AB}=0, ~~~
{\hat {\cal D}}_A \hat \rho=0,~~~
{\hat {\cal D}}_A\hat k =0,
\end{eqnarray}
that is, each level surface of $v$ is totally 
umbilic and hence spherically symmetric, which implies 
that the metric is isometric to the Reissner-Nordstrom solution.

This is of course local result since we consider only the region 
containing no saddle points of the harmonic function $v$. 
To obtain the global result, we need a further assumption such as 
analyticity. 
However, the assumption that there is no saddle point may be 
justified as follows. At a saddle point $\rho=0$, 
the level surface of $v$ is multi-sheeted, that is the embedding 
of the level surfaces is singular there. One can find at least one 
level surface such that $k_{AB}\ne 0$ near the saddle point. 
Then, Eq.~(\ref{kretchmann}) implies that the saddle point is singular. 
If the horizon is not connected, this naked singularity must  exist to 
compensate for the gravitational attraction between black holes.

\section{Dilatonic charged black holes}

We here consider the Einstein-Maxwell-dilaton system
\begin{eqnarray}
L={}^{(n)}R-2(\partial \phi)^2
-e^{-\alpha \phi}F^2,
\end{eqnarray}
for general dilaton coupling constant $\alpha>0$.

Adopting the metric form of
Eq. (\ref{metric}), we have the following equations
\begin{eqnarray}
\nabla^2V&=&C^2\frac{e^{-\alpha\phi}}{V}(\nabla \psi )^2,\\
\nabla^2 \phi&=&-{\nabla V\cdot\nabla\phi\over V}
+{\alpha\over 2}{e^{-\alpha\phi}\over V^2}(\nabla\psi)^2,\\
\nabla^2\psi&=&{\nabla V\cdot\nabla\psi\over V}+\alpha\nabla\phi\cdot\nabla\psi,
\end{eqnarray}
and
\begin{eqnarray}
R_{ij}& = &{\nabla_i\nabla_j V\over V}+2\nabla_i\phi\nabla_j\phi
-2{e^{-\alpha\phi}\over V^2}\nabla_i\psi\nabla_j\psi\nonumber\\
&&{}+{2\over n-2}{e^{-\alpha\phi}(\nabla\psi)^2\over V^2}g_{ij}.
\end{eqnarray}

Let us define the following quantities:
\begin{eqnarray}
\Phi_{\pm1}&=&\frac{1}{2}
\left[
e^{\alpha\phi/2}V
\pm{e^{-\alpha\phi/2}\over V}
-C^2(1+\lambda){e^{-\alpha\phi/2}\psi^2\over V^2}
\right],\nonumber\\
\\
\Phi_0&=&C(1+\lambda)^{1/2}{e^{-\alpha\phi/2}\psi\over V},
%\\
%%
%\Phi_1&=&\frac{1}{2}
%\left[
%e^{\alpha\phi/2}V
%+{e^{-\alpha\phi/2}\over V}
%-C^2(1+\lambda){e^{-\alpha\phi/2}\psi^2\over V^2}
%\right],\nonumber\\
\end{eqnarray}
and
\begin{eqnarray}
\Psi_{\pm 1}=\frac{1}{2}
\left(e^{-2C^2\phi/\alpha}V
\pm e^{2C^2\phi/\alpha}V^{-1}
\right),\\
%
%\Psi_{1}&=&\frac{1}{2}
%\left(e^{-2C^2\phi/\alpha}V
%+e^{2C^2\phi/\alpha}V^{-1}
%\right),
\end{eqnarray}
where
$\lambda=\alpha^2/4C^2$ has been defined.

Let us consider the conformal transformation defined by
\begin{eqnarray}
\tilde g_{ij}=V^{\frac{2}{n-3}}g_{ij},
\end{eqnarray}
and introduce the following symmetric tensors defined on this space
\begin{eqnarray}
\tilde G_{ij}=
\tilde\nabla_i\Phi_{-1}\tilde\nabla_j\Phi_{-1}
-\tilde\nabla_i\Phi_0\tilde\nabla_j\Phi_0
-\tilde\nabla_i\Phi_{1}\tilde\nabla_j\Phi_{1}
\end{eqnarray}
and
\begin{eqnarray}
\tilde H_{ij}=
\tilde\nabla_i\Psi_{-1}\tilde\nabla_j\Psi_{-1}
-\tilde\nabla_i\Psi_{1}\tilde\nabla_j\Psi_{1}.
\end{eqnarray}

Then the field equations become
\begin{eqnarray}
\tilde \nabla^2 \Phi_A &=& \tilde K \Phi_A, \label{newein1}\\
\tilde \nabla^2 \Psi_A &=&  \tilde H \Psi_A, \label{newein2}
\end{eqnarray}
and
\begin{eqnarray}
\tilde R_{ij}={2\over C^2}(1+\lambda)\left(\tilde G_{ij}+\lambda\tilde H_{ij}\right),
\label{newein3}
\end{eqnarray}
where $A=-1, 0, 1$.

Furthermore, we perform the following conformal transformations
\begin{eqnarray}
{}^{\Phi}g_{ij}^{\pm}=%{}^{\Phi}\Omega^2_{\pm} \tilde g_{ij}=
{}^{\Phi}
\omega_\pm^{2/(n-3)}\tilde g_{ij}
\end{eqnarray}
and
\begin{eqnarray}
{}^{\Psi}g_{ij}^\pm=%{}^{\Psi}\Omega^2_\pm \tilde g_{ij}=
{}^{\Psi}
\omega^{2/(n-3)}_\pm \tilde g_{ij},
\end{eqnarray}
where
\begin{eqnarray}
{}^{\Phi}\omega_\pm = \frac{\Phi_1 \pm 1}{2}
\end{eqnarray}
and
\begin{eqnarray}
{}^{\Psi}\omega_\pm = \frac{\Psi_1 \pm 1}{2}.
\end{eqnarray}
%The both of metric is related to each other as
%\begin{eqnarray}
%{}^{\Psi}g^\pm_{ij}%= \biggl( \frac{{}^{\Psi}\Omega}{{}^{\Phi}\Omega} \biggr)^2
%%{}^{\Phi}g_{ij}
%=\Omega_\pm^2{}^{\Phi}g^\pm_{ij}.
%\end{eqnarray}
%We remind the reader that we must
The extreme case should be excluded to keep 
the above conformal factors non-negative. 
In fact, there exist  multi-black hole solutions in the 
extreme limit\cite{Shiraishi}.

Now we have four manifolds 
$({}^{\Phi}\Sigma_+,{}^{\Phi}g_{ij}^+  ) $, 
$({}^{\Phi}\Sigma_-,{}^{\Phi}g_{ij}^-  ) $, 
$({}^{\Psi}\Sigma_+,{}^{\Psi}g_{ij}^+  ) $ and 
$({}^{\Psi}\Sigma_-,{}^{\Psi}g_{ij}^-  )$. Pasting 
$({}^{\Phi(\Psi)}\Sigma_\pm ,{}^{\Phi(\Psi)}g_{ij}^\pm )$ 
across the surface 
$V=0$, we can construct a complete regular surface ${}^{\Phi(\Psi)} 
\Sigma ={}^{\Phi(\Psi)}\Sigma^+ \cup {}^{\Phi(\Psi)} \Sigma^-$. 
%In the same way,
%we have another regular surface ${}^{\Psi}
%\Sigma ={}^{\Psi}\Sigma^+ \cup {}^{\Psi} \Sigma^-$. 
Thus, we 
have two regular surfaces, ${}^{\Phi}\Sigma$ and ${}^{\Psi}\Sigma$. 
As in the 
previous section, we can check that each total gravitational 
mass on ${}^{\Phi}\Sigma$ and ${}^{\Psi}\Sigma$ vanishes.

From now on we use the conformal positive energy 
theorem\cite{CPET} to show that the static slice is conformally flat. 
See Appendix A for the conformal positive energy theorem in higher dimensions. 
We consider another conformal transformation 
\begin{equation}
\hat g^\pm_{ij}:=\left[
({}^\Phi\omega_\pm)^2({}^\Psi\omega_\pm)^{2\lambda}\right]^{{1\over(n-3)(1+\lambda)}}\tilde g_{ij}
\end{equation}
The Ricci curvature on this space can be shown to become
\begin{eqnarray}
(1+\lambda) \hat R^\pm & = &
\left[
({}^\Phi\omega_\pm)^2({}^\Psi\omega_\pm)^{2\lambda}\right]^{-{1\over(n-3)(1+\lambda)}}
\nonumber\\
&&\times\left[
({}^\Phi\omega_\pm)^{{2\over n-3}}({}^\Phi R^\pm)
+\lambda({}^\Psi\omega_\pm)^{{2\over n-3}}({}^\Psi R^\pm)\right]
\nonumber\\
&&{}+{(n-2)\lambda\over 1+\lambda}
\left(\hat\nabla\ln{}^\Psi\omega_\pm
-\hat\nabla\ln{}^\Phi\omega_\pm\right)^2.
\end{eqnarray}
The first term of the r.h.s. turns out to be non-negative:
\begin{eqnarray}
&&({}^\Phi\omega_\pm)^{{2\over n-3}}({}^\Phi R^\pm)
+\lambda({}^\Psi\omega_\pm)^{{2\over n-3}}({}^\Psi R^\pm)
\nonumber\\
&&={2\over C^2}\Biggl|
{\Phi_0\tilde\nabla\Phi_{-1}-\Phi_{-1}\tilde\nabla\Phi_0
\over \Phi_1\pm 1}
\Biggr|^2.
\end{eqnarray}
%When we moved from the first to second line, we used Eq (\ref{newein3}). 
The conformal positive mass theorem implies that
\begin{equation}
\frac{{}^\Phi\omega_\pm}{{}^\Psi\omega_\pm}={\rm const.},
\end{equation}
\begin{equation}
\Phi_0={\rm const.}\times \Phi_{-1}
\end{equation}
and that each $ ({}^{\Phi}\Sigma,{}^{\Phi}g_{ij}  ) $,
$({}^{\Psi}\Sigma,{}^{\Psi}g_{ij} )$ and $(\hat \Sigma,
\hat g_{ij})$ is flat space. 
In other words,  $(\Sigma, g_{ij})$ is conformally flat.
%Using the above results
%we can show that the function $\tau$ defined by 
We define the function
\begin{equation}
v := ({}^{\Phi}\omega_\pm V )^{-1/2}.
\end{equation}
Noting that
\begin{equation}
\hat g_{ij}= v^{\frac{4}{n-3}}{}^{\Phi}g_{ij},
\end{equation}
we have
\begin{equation}
v^{4/(n-3)} \hat R = {}^{\Phi}R-\frac{4(n-2)}{n-3}
 \frac{\nabla_0^2 v}{v}.
\end{equation}
Since we already knows that 
$\hat R =  {}^{\Phi}R=0$, 
$v$ turns out to be the harmonic function on the flat space:
\begin{equation}
\nabla_0^2 v =0. \label{harmonicf}
\end{equation}

With the  procedure given in the previous section, we can show that the 
static solution must hyperspherically symmetric, 
therefore given by the metric given in Ref.\cite{GM}

\section{Branes}

As already mentioned, one motivation for the present work 
was to establish  the uniqueness of static 
$p$-brane solutions. In general these take the form
\begin{equation}
ds^2 = e^{2\gamma \Phi} ( d{\bf y}_p ^2 ) + e^ {2 \delta \Phi} g_{\mu 
\nu} dx^\mu dx ^\nu. \label{metric5}
\end{equation}
Dimensional reduction on ${\bf E} ^p$ will take the Einstein-Hilbert 
action to the Einstein-Hilbert action if $(n-2) \delta +p \gamma =0$. 
If the higher dimensional metric is coupled to an $n-2$ form $F_{n-2}$ 
with no scalars, we obtain the Lagrangian\cite{GHT}
\begin{equation}
R -2 (\nabla \phi)^2 - { 2 \over (n-2)! } e^{ -2 \alpha \phi} F^2 _{n-2},
\end{equation}
with
\begin{equation}
\gamma^2 = { 2 (n-2) \over p(n+p-2) }
\end{equation}
and
\begin{equation}
\alpha^2 = { 2 p(n-3)^2 \over (n-2) (n+p-2) }.
\end{equation}
In this way, we obtain  a uniqueness theorem for black $p$-branes 
described by the metric ansatz of Eq (\ref{metric5}). Note that, 
by contrast with our previous paper \cite{Ida2}, the values of 
$(n,p)$ are unrestricted.

\section{Summary}

We presented the proof of the uniqueness theorem for  static 
charged dilatonic black holes in higher dimensions. We excluded 
the extreme case from consideration. Our theorem also provides a 
uniqueness theorem for black $p$-branes.

Since the extreme case is a BPS state, the remaining issue 
about the extreme case is  important. 
In this paper we considered only electrically charged black holes. 
However, the generalization to the magnetically charged case is 
straightforward because of  the duality as mentioned in the last section.

Finally we comment on the assumption on asymptotic flatness. 
If we drop, there will be infinite sequence  of  solutions 
constructed from the  the Bohm metrics \cite{Bohm} as in 
our previous work \cite{Ida1,Ida2}. 
Since the positive energy theorem does not hold in such cases, 
it is unclear whether proof of the uniqueness presented here can 
be generalized to cover this case. The stability of Bohm black 
holes  solutions is currently under investigation.

\vskip 1cm

{\em Acknowledgments.}

TS's work is partially supported by Yamada Science Foundation and 
Grant-in-Aid for Scientific Research from Ministry of Education, Science, 
Sports and Culture of Japan(No. 13135208, No.14740155 and No.14102004).

\appendix

\section{The conformal positive energy theorem in higher dimensions}

In this appendix we briefly describe the statement and the proof of 
the conformal positive energy theorem in higher dimensions. 
The proof is given by a straightforward  extension of that of 
Simon \cite{CPET} in four dimensions. 
Simon in turn was inspired by Masood-ul-Alam's proof of the 
uniqueness of the Gibbons-Maeda solution in four dimensions.

{\it Theorem:}Let $({}^{\Phi}\Sigma,{}^{\Phi}g_{ij})$ and 
$({}^{\Psi}\Sigma,{}^{\Psi}g_{ij})$ be asymptotically flat 
Riemannian $(n-1)$-dimensional manifolds with ${}^{\Psi}g_{ij} 
=\Omega^2 {}^{\Phi}g_{ij} $. Then ${}^{\Phi}m+\beta {}^{\Psi}m \geq 0 $ 
if ${}^{\Phi}R+\beta \Omega^2 {}^{\Psi}R \geq 0  $ hold for 
a positive constant $\beta$. The equality 
holds if and only if $({}^{\Phi}\Sigma,{}^{\Phi}g_{ij})$ and 
$({}^{\Psi}\Sigma,{}^{\Psi}g_{ij})$ are flat.

{\it Proof:} Let us consider the conformal transformation as 
$\overline g_{ij} =\Omega^{\frac{2\beta}{1+\beta}}{}^{\Phi}g_{ij}$. 
It is easy to see that
\begin{eqnarray}
(1+\beta)\overline R & = &  \Omega^{-\frac{2\beta}{1+\beta}}
({}^{\Phi}R + \beta \Omega^2 {}^{\Psi}R) \nonumber \\
& & +(n-2)(n-3)\frac{\beta}{1+\beta}
\biggl( \frac{\overline D \Omega }{\Omega} \biggr)^2.
\end{eqnarray}
Like Witten's positive energy theorem\cite{PET,Marika}, now, 
we can prove the positivity of the total 
gravitational mass on $(\overline \Sigma, \overline g_{ij})$ 
which is
\begin{eqnarray}
\overline m = (1+\beta)^{-1}({}^{\Phi}m+ \beta {}^{\Psi}m) \geq 0.
\end{eqnarray}

For the case of $\overline m =0$, we see that  $(\overline \Sigma, 
\overline g_{ij})$ is flat and $\Omega$ is constant. They imply that 
$({}^{\Phi}\Sigma,{}^{\Phi}g_{ij})$ and $({}^{\Psi}\Sigma,{}^{\Psi}g_{ij})$ 
are also flat.

It is trivial that $\overline m =0 $ if $({}^{\Phi}\Sigma,{}^{\Phi}g_{ij})$ 
and $({}^{\Psi}\Sigma,{}^{\Psi}g_{ij})$ are flat. {\it Q.E.D.}

\end{document}